\documentclass[letterpaper, 10 pt, conference]{ieeeconf}  

\IEEEoverridecommandlockouts                              
\usepackage{graphics} 
\usepackage{epsfig} 
\usepackage{mathptmx} 
\usepackage{times} 
\usepackage{amsmath} 
\usepackage{amssymb}  
\usepackage[monochrome]{xcolor} 
\usepackage{algorithm}
\usepackage{algorithmic}
\usepackage{wrapfig}
\usepackage[hidelinks]{hyperref}
\usepackage{multirow} 
\usepackage{booktabs} 
\newcommand{\D}[1]{\textcolor{blue}{#1}}

\title{\LARGE \bf
Learning to Control PDEs with Differentiable Predictive Control and Time-Integrated Neural Operators
}

\author{Dibakar Roy Sarkar, J\'an Drgo\v{n}a and Somdatta Goswami
\thanks{Authors are with the Department of Civil and Systems Engineering,
Johns Hopkins University, Baltimore, MD 21218 USA.}
\thanks{ The research efforts of DRS and SG are supported by the National Science Foundation (NSF) under Grant No. 2436738. JD was supported by the US DOE, Office of Science, ASCR program under the Scientific Discovery through Advanced Computing (SciDAC) Institute “LEADS: LEarning-Accelerated Domain Science”. 
We acknowledge the computing support provided by the Advanced Research Computing at Hopkins (ARCH) core facility at Johns Hopkins University and the Rockfish cluster. }
}

\begin{document}

\maketitle
\thispagestyle{empty}
\pagestyle{empty}

\begin{abstract}
We present a data-driven control framework for partial differential equations (PDEs). 
Our approach integrates time-integrated deep operator networks (TI-DeepONet) as differentiable PDE surrogate models within  differentiable predictive control (DPC)---a self-supervised learning framework for constrained neural control policies.
The TI-DeepONet architecture learns temporal derivatives and couples them with numerical integrators, while the DPC algorithm uses automatic differentiation to compute policy gradients by backpropagating the expectations of the optimal control loss through the learned TI-DeepONet. This approach enables efficient offline optimization of neural policies without the need for online optimization or supervisory controllers.
We empirically demonstrate the proposed method across diverse PDE systems, including the heat, the nonlinear inviscid Burgers’, and the reaction-diffusion equations. The learned policies achieve target tracking, constraint satisfaction, and curvature minimization objectives, while generalizing across distributions of initial conditions and  parameters. 
Moreover, we demonstrate four orders of magnitude acceleration at inference time compared to nonlinear model predictive control benchmarks.
These results highlight the promise of neural operator learning for scalable model-based control of PDEs.

\end{abstract}

\section{INTRODUCTION}

Control of physical systems governed by partial differential equations (PDEs) is central to engineering applications ranging from energy systems to biomedical devices~\cite{Bhan2024, krstic2008boundary, meurer2009tracking}. In these problems, the controller must steer a spatiotemporal field toward a desired behavior through distributed or boundary inputs, while satisfying physical dynamics, safety constraints, and actuator bounds in real time. This task is made particularly demanding by the infinite-dimensional nature of PDE state spaces and, in many practical settings, the unavailability of explicit governing equations, making the design of computationally tractable, data-driven controllers a central open challenge.

A natural starting point is model predictive control (MPC), which handles constraints and multi-step objectives by solving an optimization problem at each time step. On PDE-driven systems, however, the cost of running a high-fidelity physics-based solver at every step is computationally prohibitive as spatial resolution or prediction horizons grow. One strategy is to replace the PDE solver with a learned surrogate inside the MPC loop. Surrogate methods based on Koopman operators~\cite{korda2018linear} or neural operators~\cite{sirota2025pde} reduce per-step cost while preserving constraint handling. More directly, \cite{de2025deep} proposes a deep operator network (DeepONet)-based MPC framework for real-time PDE control, but still requires solving an optimization problem at every time step. Similarly, \cite{pmlr-v283-li25d} accelerates MPC for nonlinear PDE systems using physics-informed neural surrogates, but assumes the governing PDE is fully known and available for training.

A structurally different approach is analytical backstepping~\cite{krstic2008boundary, vazquez2008control}, which constructs explicit stabilizing controllers through Volterra-type coordinate transformations. Computing the resulting control kernels requires solving complex integro-differential equations~\cite{krstic2024machine}, and neural operators have been applied to learn these kernel gain mappings~\cite{qi2024neural, bhan2023neural}, reducing computational cost while retaining the backstepping structure. Most directly, \cite{bhan2023neural} learns the backstepping kernel operator via neural operators, but still requires full knowledge of the governing PDE and is designed for stabilization rather than general tracking or constraint satisfaction.

To avoid online optimization altogether, a parallel line of work learns control policies offline. \cite{Holl2020Learning} leverages differentiable PDE solvers with the adjoint method to compute policy gradients from physical simulations, while \cite{Hwang_Lee_Shin_Hwang_2022} employs an autoencoder-based surrogate within a gradient-based open-loop optimization framework. Imitation learning offers a related strategy, where a neural network approximates an MPC policy from offline-generated labels~\cite{hertneck2018learning, chen2018approximating, karg2020efficient}, enabling real-time execution but still requiring the expensive MPC problem to be solved offline for data generation. Reinforcement learning offers yet another alternative through direct interaction with PDE dynamics~\cite{rabault2019artificial, bucci2019control}, though it faces challenges in sample efficiency and scalability to high-dimensional state spaces.  \D{More recent work augments a neural operator with a physics-informed loss and an online adaptation scheme~\cite{lundqvist2025residual}.}
Most closely related, \cite{guven2025learning} jointly trains a neural operator for dynamics and a neural policy for control, but assumes explicit knowledge of the PDE and couples both objectives in a single training loop, causing conflicting gradients, while \cite{zanotta2026cinoc} trains operator control policies by differentiating through a numerical PDE solver. In contrast, our framework decouples dynamics learning from policy synthesis and requires neither prior knowledge of the governing PDE nor a solver in the training loop.

Differentiable predictive control (DPC)~\cite{drgovna2024learning, DRGONA202280} offers a principled path in this direction. Rather than solving an optimization problem online, DPC learns an explicit policy offline by backpropagating the MPC objective through a differentiable dynamics model, eliminating the need for expert demonstrations or online solvers while naturally handling constraints. DPC has been demonstrated for ODE systems, but extending it to PDEs requires surrogate models that are both differentiable and accurate over long prediction horizons. Standard architectures such as DeepONet~\cite{lu2021learning} and FNO~\cite{kovachki2021neural} can approximate PDE solution operators, yet autoregressive rollouts accumulate errors over long horizons and both face challenges in temporal extrapolation~\cite{wang2025time,michalowska2025multi}. We therefore adopt time-integrated DeepONet (TI-DeepONet)~\cite{NAYAK2026118960} as our surrogate of choice, which learns the temporal derivative operator and couples it with classical numerical integrators, preserving temporal causality and yielding stable gradients over long horizons.

\paragraph{Contributions}
This work makes the following contributions: (1) End-to-end differentiable learning-to-control: we integrate TI-DeepONet with DPC to enable differentiable closed-loop PDE control; (2) Scalable offline policy learning: we learn constrained neural policies offline, eliminating online optimization and supervisory controllers while generalizing across initial conditions and PDE parameters; (3) Validation on canonical PDEs: we demonstrate accurate target tracking, shock mitigation, and constraint satisfaction across three representative systems; and (4) Open-source implementation: we release the code and trained models to support reproducibility and further research.\footnote{\url{https://github.com/Centrum-IntelliPhysics/PDEControl_DPC}}

\section{Problem Formulation}

Consider a general time-dependent PDE describing the evolution of a spatiotemporal field $u(\mathbf{x},t)$ over a spatial domain $\mathbf{x}\in\Omega\subset\mathbb{R}^d$ and time horizon $t\in[0,T]$, where $d$ is the spatial dimension and $T>0$ is the prescribed time horizon:
\begin{equation}
    \frac{\partial u}{\partial t}(\mathbf{x}, t) 
    = \mathcal{F}\big(t, \mathbf{x}, u, \nabla u, \nabla^2 u, \ldots, a(\mathbf{x}, t)\big),
    \label{eq:pde}
\end{equation}
subject to the initial condition $u(\mathbf{x}, 0) = u_0(\mathbf{x})$ 
and boundary conditions $\mathcal{B}(u) = 0$ on $\partial \Omega$. 
Here, $\mathcal{F}$ denotes a (possibly nonlinear) differential operator, and 
$a(\mathbf{x}, t) \in \mathcal{A} \subseteq \mathbb{R}^{n_a}$ 
is the distributed control input acting over the domain $\Omega$. The continuous-time PDE-constrained optimal control problem seeks the optimal control function 
$a^*(\mathbf{x}, t)$ that minimizes an objective functional of the form:
\begin{subequations}\label{eq:ocp_cont}
\begin{align}
    \min_{a(\cdot,\cdot)} \quad
    & J[u, a] = \int_0^T \!\! \int_{\Omega} 
    \ell\!\left(u(t,\mathbf{x}), a(t,\mathbf{x}), \boldsymbol{\xi}(t)\right) 
    \, d\mathbf{x}\, dt \\
    &+ \int_{\Omega} \ell_T\!\left(u(T,\mathbf{x})\right) d\mathbf{x}, 
    \label{eq:ocp_cost}\\
    \text{s.t.} \quad
    & \frac{\partial u}{\partial t} 
    = \mathcal{F}(t, \mathbf{x}, u, \nabla u, \nabla^2 u, \ldots, a), 
    \label{eq:ocp_dyn}\\
    & h(u(\mathbf{x}, t), \boldsymbol{\xi}(t)) \le 0, \quad
      g(a(\mathbf{x}, t), \boldsymbol{\xi}(t)) \le 0, 
    \label{eq:ocp_cons}\\
    & u(\mathbf{x}, 0) = u_0(\mathbf{x}), \quad 
      \mathcal{B}(u) = 0. 
    \label{eq:ocp_ic_bc}
\end{align}
\end{subequations}
Here, $\ell$ and $\ell_T$ denote the running and terminal cost densities; 
$h$ and $g$ encode spatially distributed state and control constraints;  and $\boldsymbol{\xi}(t)$ represents time-varying parameters such as reference trajectories,  physical coefficients, or constraint bounds. \D{Here, $a^*(\mathbf{x}, t)$ denotes the optimal control for given initial condition $u_0(\mathbf{x})$ and parameters $\boldsymbol{\xi}(t)$.}

\section{Methodology}

Solving~\eqref{eq:ocp_cont} directly is computationally prohibitive due to the infinite-dimensional  state and control spaces. 
Traditional methods rely on spatial discretization and online numerical optimization 
(e.g., PDE-MPC). However, discretize-then-optimize approach scales poorly for high-dimensional domains or long horizons.
To overcome these challenges, we introduce an end-to-end differentiable surrogate-based formulation 
in which both the PDE dynamics and control policy are approximated by neural representations, while 
enabling efficient offline gradient-based policy optimization.

\begin{figure}[!tb]
    \centering
    \includegraphics[width=0.95\linewidth]{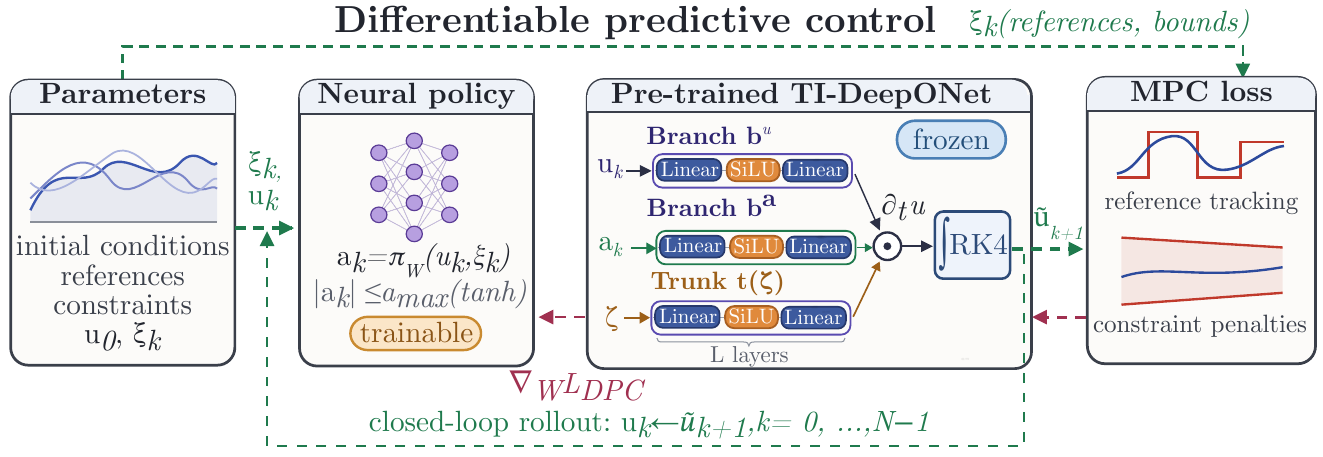}
    \caption{\D{Schematic of the proposed TI-DeepONet–DPC architecture. At each step $k$, the neural policy $\pi_{\mathbf W}$ maps the state $\mathbf u_k$ and problem parameters $\boldsymbol\xi_k$ to control $\mathbf a_k$. The pre-trained TI-DeepONet $\mathcal G_\theta$ predicts $\partial_t\mathbf u$, which is integrated via RK4 to obtain $\mathbf u_{k+1}$. Forward propagation (green dashed arrows) rolls out the closed-loop dynamics, while backward propagation (red dashed arrows) differentiates the MPC loss \eqref{eq:dpc_loss} through the rollout to compute $\nabla_{\mathbf W}\mathcal L_{\mathrm{DPC}}$, enabling offline learning of constrained PDE control policies.}}
    \label{fig:DPC_NO_schematic}
\end{figure}

Specifically, we propose integration of \emph{TI-DeepONet} with \emph{DPC} to learn neural control policies for PDE-constrained parametric optimal control problems. Our methodology is conceptually illustrated in Figure~\ref{fig:DPC_NO_schematic}.
The TI-DeepONet component serves as a differentiable surrogate for the PDE dynamics, 
capturing the instantaneous temporal derivatives of the governing equations and integrating them 
through numerical solvers to generate finite-horizon rollouts. 
The DPC algorithm then enables scalable offline computation of the policy gradients via backpropagation of the expectation of the control loss through the unrolled PDE dynamics.
This combination yields a fully differentiable closed-loop system that maintains physical consistency, 
supports constraint handling, and eliminates the need for online optimization.

\subsection{Time-integrated deep operator network (TI-DeepONet)}

Standard DeepONets suffer from error accumulation during autoregressive time-stepping, limiting long-horizon predictions. TI-DeepONet~\cite{NAYAK2026118960} mitigates this by learning the temporal derivative $\partial u/\partial t \approx \mathcal{G}\theta(t,\mathbf{x},u,\nabla u,\nabla^2u,\ldots,a)$ rather than the state directly, which is then integrated using classical schemes such as fourth-order Runge--Kutta (RK4). This preserves the Markovian structure of the PDE dynamics, with each prediction depending only on the current state. For numerical PDE control, we discretize the spatial domain $\Omega$ using a uniform grid with $n_x$ points $x_1,\ldots,x{n_x}$. The continuous fields $u(t,\mathbf{x})$ and $a(t,\mathbf{x})$ are then represented by their discrete values $u_i(t)$ and $a_i(t)$ at each grid point. For PDE control, we modify this architecture with two branch networks: a state branch $\mathbf{b}^u$ encoding the solution field $u^i$, and a control branch $\mathbf{b}^a$ encoding the control function $a^i$. These are combined as:
\begin{equation}
    \partial u/\partial t  \approx\mathcal{G}_\theta(u^i, a^i)(\zeta) = \sum_{j=1}^{p} \left[b^u_j(u^i) \odot b^a_j(a^i)\right] t_j(\zeta)
    \label{eq:dual_branch}
\end{equation}
where $\odot$ denotes element-wise multiplication, capturing the coupled influence of state and control on temporal evolution. \D{Here, $b^u_j(\cdot)$ and $b^a_j(\cdot)$ are the $j$-th outputs of the state and control branch networks, $t_j(\zeta)$ is the $j$-th output of the trunk network evaluated at the spatial query coordinate $\zeta \in \Omega$, and $p$ is the number of latent basis functions. We refer the reader to~\cite{NAYAK2026118960} for detailed architectural specifications.}

\subsection{Differentiable Predictive Control with TI-DeepONet}

Given the differentiable dynamics model $\mathcal{G}_{\boldsymbol{\theta}}$, 
we formulate a finite-dimensional discrete-time approximation of~\eqref{eq:ocp_cont} over a horizon $N$ as the following 
parametric optimal control problem:
\begin{subequations}\label{eq:pOCP}
\begin{align}
    \min_{\mathbf{W}} ~
    & \mathbb{E}_{u_0 \sim \mathcal{P}_{u_0}, \boldsymbol{\xi} \sim \mathcal{P}_{\boldsymbol{\xi}}}
    \!\left[\sum_{k=0}^{N-1} \sum_{i=1}^{n_x} \ell(u_k^i, a_k^i, \boldsymbol{\xi}_k) \Delta x
    + \sum_{i=1}^{n_x} \ell_N(u_N^i) \Delta x \right], \label{eq:pOCP_obj}\\
    \text{s.t.} \quad 
    & \mathbf{u}_{k+1} = \mathrm{ODESolve}\!\left(\mathcal{G}_{\boldsymbol{\theta}}(t_k, \mathbf{u}_k, \mathbf{a}_k)\right), \\
    & \mathbf{a}_k = \pi_{\mathbf{W}}(\mathbf{u}_k, \boldsymbol{\xi}_k), ~ 
      h(\mathbf{u}_k,\boldsymbol{\xi}_k) \le 0, ~ g(\mathbf{a}_k,\boldsymbol{\xi}_k) \le 0,
\end{align}
\end{subequations}
Here, $\pi_{\mathbf{W}}$ is a neural policy parameterized by weights $\mathbf{W}$; $\ell(\cdot)$ and $\ell_N(\cdot)$ represent the stage cost and terminal cost functions, respectively; $h(\cdot)$ and $g(\cdot)$ encode state and control constraints; and $\boldsymbol{\xi}_k$ contains time-varying problem parameters such as reference signals, state constraints, and control constraints. The expectation in \eqref{eq:pOCP_obj} is taken over distributions of initial conditions $\mathcal{P}_{u_0}$ and problem parameters $\mathcal{P}_{\boldsymbol{\xi}}$, enabling the learned policy to generalize across multiple scenarios.

\D{Unlike MPC, which relies on discretization and online optimization, DPC optimizes the policy parameters $\mathbf{W}$ offline to minimize the expected cost over a distribution of initial conditions and parameters, yielding an explicit control policy that can be evaluated rapidly online without iterative optimization.}

\paragraph{Neural Control Policy Architecture.}
The control policy $\pi_{\mathbf{W}}: \mathcal{U} \times \mathcal{P} \to \mathcal{A}$ is parameterized as an $L$-layer MLP with SiLU activations, ($\sigma(x) = x \cdot \mathrm{sigmoid}(x)$)\D{, in the hidden layers $\mathbf{z}_{\ell+1} = \sigma(\mathbf{H}_\ell \mathbf{z}_\ell + \mathbf{b}_\ell)$, $\ell = 0, \ldots, L-1$, mapping}  the concatenated input $\mathbf{z}_0 = [\mathbf{u}_k;\, \boldsymbol{\xi}_k] \in \mathbb{R}^{n_u + n_\xi}$ to actions $a_k \in \mathbb{R}^{n_a}$ \D{via the output layer:}
\begin{equation}
    \pi_{\mathbf{W}}(\mathbf{u}_k, \boldsymbol{\xi}_k) = a_{\max} \cdot \tanh(\mathbf{H}_L \mathbf{z}_L + \mathbf{b}_L), \label{eq:policy_output}
\end{equation}
where $\mathbf{W} = \{\mathbf{H}_\ell, \mathbf{b}_\ell\}_{\ell=0}^L$ collects all network weights and biases, with $\mathbf{H}_\ell \in \mathbb{R}^{n_{\ell+1} \times n_\ell}$ and $\mathbf{b}_\ell \in \mathbb{R}^{n_{\ell+1}}$, and $\mathbf{z}_L \in \mathbb{R}^{n_L}$ is the output of the final hidden layer. The $\tanh$ saturation\D{, applied only at the output layer} in \eqref{eq:policy_output}, enforces hard control constraints $|a_k| \le a_{\max}$ by construction, ensuring feasible actions without penalty-based soft constraint handling.

\begin{minipage}[t]{0.46\textwidth}
\begin{algorithm}[H]
\small
\caption{Train TI-DeepONet}
\begin{algorithmic}[1]
\label{algo:algorithm1}
\STATE Training datasets of initial conditions, problem parameters $\xi_k$ and control inputs from $\mathcal{P}_{u_0}$, $\mathcal{P}_{\xi}$ and $\mathcal{P}_{a}$, respectively.
\STATE Labels: $u(t)$, $f(t)$ $\to$ $u(t+1)$
\STATE Neural operator, $\mathcal{G}_{\theta}: (u(t), a(t)) \mapsto \partial_t u$
\STATE Time stepping, $u(t+1) = \text{RK}4(\mathcal{G}_{\theta})$
\STATE Mean square error loss, $\mathcal{L}_{\text{mse}}$
\STATE Optimizer $\mathbb{O}$
\STATE Learn, $\theta$ via optimizer $\mathbb{O}$ using gradient $\nabla_{\theta}\mathcal{L}_{\text{mse}}$
\RETURN trained neural operator, $\mathcal{G}_{\theta}$
\end{algorithmic}
\end{algorithm}
\end{minipage}

\begin{minipage}[t]{0.46\textwidth}
\begin{algorithm}[H]
\small
\caption{DPC offline training}
\begin{algorithmic}[1]
\label{algo:algorithm2}
\STATE Training datasets of initial condition and problem parameters $\xi_k$ from $\mathcal{P}_{u_0}$ and $\mathcal{P}_{\xi}$ respectively.
\STATE Pretrained neural operator, $\mathcal{G}_{\theta}$
\STATE Time stepping, $u(t+1) = \text{RK}4(\mathcal{G}_{\theta})$
\D{\STATE DPC loss, $\mathcal{L}_{\text{DPC}}$ \eqref{eq:dpc_loss}}
\STATE Optimizer $\mathbb{O}$
\STATE Learn, $\mathbf{W}$ via optimizer $\mathbb{O}$ using gradient $\nabla_{\mathbf{W}}\mathcal{L}_{\text{DPC}}$
\RETURN optimized policy $\pi_{\mathbf{W}}(\mathbf{u}_k, \boldsymbol{\xi}_k)$
\end{algorithmic}
\end{algorithm}
\end{minipage}
\hspace{0.2cm}

\paragraph{Loss Function and Policy Gradients.}
\D{To enable offline policy learning, we relax~\eqref{eq:pOCP} into an empirical risk minimization loss with constraints handled via penalties, extending the standard DPC formulation~\cite{drgovna2024learning, DRGONA202280} to the spatiotemporal setting by averaging over a batch of $m$ sampled trajectories, each rolled out over $N$ time steps across $n_x$ spatial points:
\begin{multline}
\mathcal{L}_{\text{DPC}} = \frac{1}{m}\sum_{s=1}^{m}\sum_{i=1}^{n_x}\Delta x \Bigg[ Q_N\, \ell_N\big(u_N^{i,s}\big) + \sum_{k=0}^{N-1}\Big( Q_\ell\, \ell\big(u_k^{i,s}, a_k^{i,s}, \boldsymbol{\xi}_k^{s}\big) \\
+ Q_h \big\|\text{ReLU}\big(h(u_k^{i,s}, \boldsymbol{\xi}_k^{s})\big)\big\|^2_2 + Q_g \big\|\text{ReLU}\big(g(a_k^{i,s}, \boldsymbol{\xi}_k^{s})\big)\big\|^2_2 \Big)\Bigg],
\label{eq:dpc_loss}
\end{multline}
where superscript $s$ indexes sampled scenarios, $\text{ReLU}(\cdot)$ ensures that only violated constraints contribute to the loss, and the weights $Q_\ell, Q_h, Q_g, Q_N \in \mathbb{R}_{>0}$ balance the relative importance of performance and feasibility.}
 
 The expectation over initial conditions $u_0 \sim \mathcal{P}_{u_0}$ and problem parameters $\boldsymbol{\xi}_k \sim \mathcal{P}_{\boldsymbol{\xi}}$ is approximated via mini-batch sampling, enabling the learned policy to generalize across diverse scenarios. Since both TI-DeepONet surrogate $\mathcal{G}_\theta$ and the neural policy $\pi_{\mathbf{W}}$ are differentiable, \D{the gradient of \eqref{eq:dpc_loss} flows from the loss through the rolled-out state trajectory and the TI-DeepONet dynamics into the policy network, yielding $\nabla_{\mathbf{W}} \mathcal{L}_{\text{DPC}}$ end-to-end via automatic differentiation. This enables direct policy optimization with standard gradient-based methods, without an auxiliary expert controller providing demonstrations or reference trajectories.}

\paragraph{Training procedure.} We employ a two-stage approach: first as defined in~Algorithm~\ref{algo:algorithm1}, $\mathcal{G}_\theta$ is pretrained on simulation data to learn PDE dynamics; second as defined in~Algorithm~\ref{algo:algorithm2}, the operator is frozen and policy parameters $W$ are optimized via stochastic gradient descent on $\mathcal{L}_{\text{DPC}}$ over sampled initial conditions $u_0 \sim \mathcal{P}_{u_0}$ and problem parameters $\xi_k \sim \mathcal{P}_\xi$.

\section{Experiments}

We evaluate our approach on three canonical 1D PDEs representing distinct dynamical behaviors: the heat equation (parabolic diffusion), Burgers' equation (hyperbolic shock formation), and the Fisher-KPP equation (reaction-diffusion). \D{To benchmark the learned policies, we compare against a nonlinear model predictive control (NMPC) baseline~\cite{kouvaritakis2016model} that, at each sampling instant, solves a finite-horizon discretization of~\eqref{eq:ocp_cont} via direct multiple shooting, with the finite difference method (FDM)-discretized PDE dynamics imposed as equality constraints and control bounds as box constraints. The resulting nonlinear program is formulated in CasADi and solved with the interior-point solver IPOPT with warm starting; only the first control input is applied in a receding-horizon fashion.} Our experiments demonstrate that the learned control policies can stabilize systems, track targets, and satisfy constraints using a differentiable surrogate model.

\subsection{Problem Setup and Control Parametrization}

\noindent\textbf{Control Representation.} All control inputs are parameterized via Gaussian basis functions:
\begin{equation}
    f(x,t) = \sum_{i=1}^{n} f_i(t) \cdot \exp\left(-\frac{(x-\mu_i)^2}{2\sigma^2}\right),
    \label{eq:control_basis}
\end{equation}
where $n$ is the number of actuators, $\mu_i$ denotes the $i$-th actuator location, $f_i(t)$ are learned time-varying control amplitudes, and $\sigma$ controls the spatial influence of each actuator.

\noindent \textbf{Numerical Implementation.} We employ spatial discretization $\Delta x = 10^{-2}$ and temporal discretization $\Delta t = 10^{-3}$ for all solvers. Control amplitudes $f_i(t)$ are generated from perturbed sinusoidal signals with problem-specific bounds.

\noindent \textbf{Learning Framework.} For each system, we train a TI-DeepONet $\mathcal{G}_{\theta}: (u(t), f(t)) \mapsto \partial_t u$. The learned operator is then integrated using RK4 for time evolution during training, enabling DPC by solving \eqref{eq:pOCP}. 

\subsection{Heat Equation (HE): Thermal Target Tracking} 
\label{sec:heat}

\begin{figure}[!bt]
    \centering
    \includegraphics[width=0.5\textwidth]{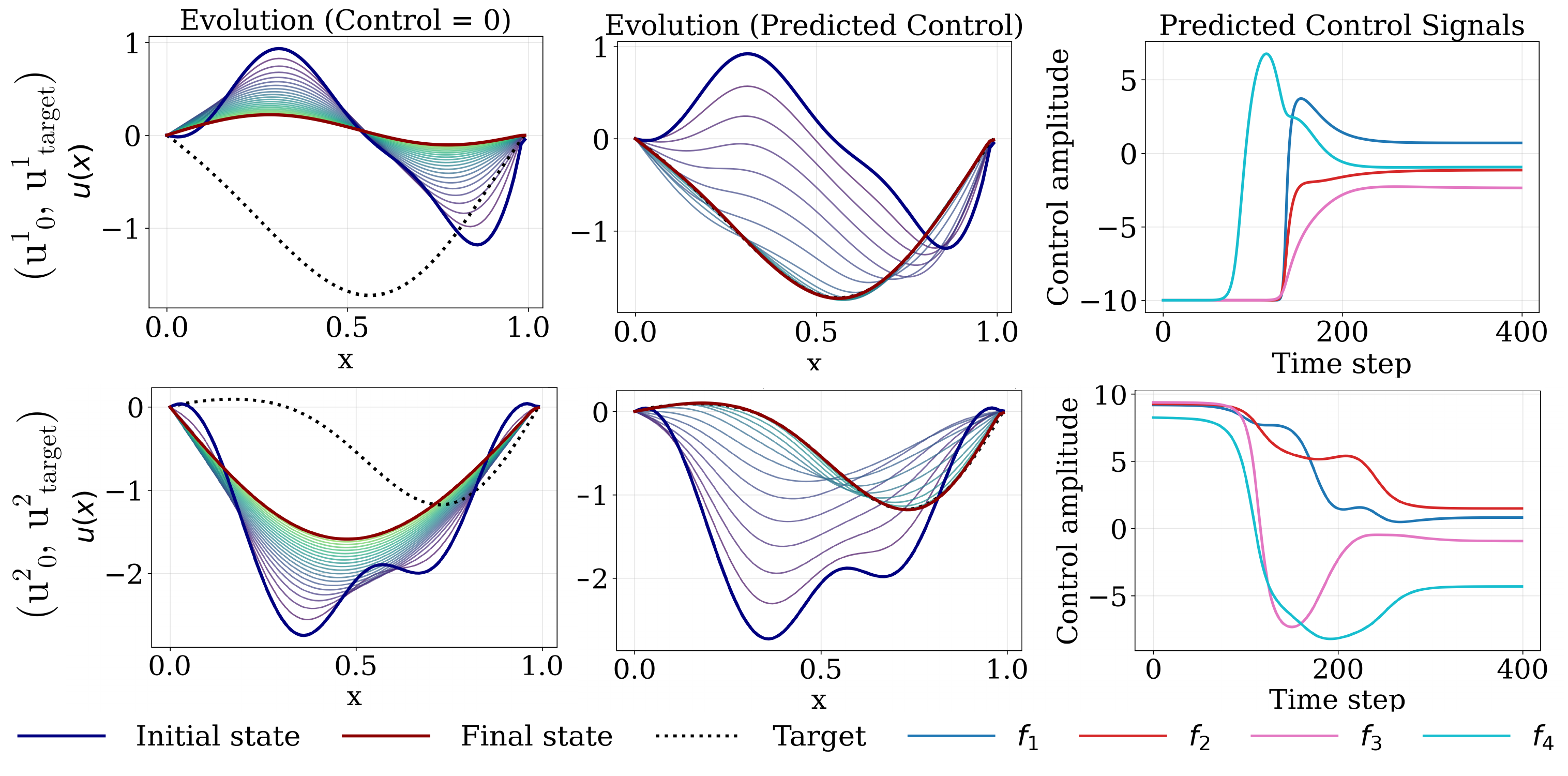}
        \vspace{-0.6cm}
    \caption{\D{HE control performance. Here $u_0^i$ denotes the initial state and $u_{\mathrm{target}}^i$ the target state for the $i$-th scenario. Each row presents a scenario showing: \textit{(left)} uncontrolled evolution from initial state (blue) to final state (red) versus target (black dotted); \textit{(middle)} controlled trajectory achieving the target; \textit{(right)} applied control signals $f_i(t)$.}}
    \label{fig:heat1D_DPC_results}
\end{figure}

\noindent\textbf{System Dynamics.} Consider the HE with Dirichlet boundary conditions:
\begin{equation}
\frac{\partial u}{\partial t} = \alpha \frac{\partial^2 u}{\partial x^2} + f(x,t), ~~ u(0,t) = u(1,t) = 0,
\end{equation}
where $x \in [0,1], \, t \in [0,T]$. $u(x,t)$ is the temperature field, $\alpha = 0.1$ is the thermal diffusivity. Control is applied via $n = 4$ actuators positioned at $\mu_i \in \{0.2, 0.4, 0.6, 0.8\}$ with spread $\sigma = 0.1$, and terminal time $T = 0.4$.

\noindent \textbf{Data Generation and Operator Learning.} We employ the FDM with the unconditionally stable Crank-Nicolson scheme. Training data comprises 3000 samples with initial conditions drawn from a Gaussian random field (GRF) using an RBF kernel (length scale $l = 0.4$, variance $\sigma_u^2 = 4$). Control amplitudes are bounded: $|f_i(t)| \leq \D{10}$. The trained TI-DeepONet achieves average relative $L_2$ error of $(1.64 \pm 1.13) \times 10^{-2}$ on 3000 test rollouts.

\noindent \textbf{Control Objective.} Given a feasible target temperature profile $u_{\text{target}}$, the goal is to minimize the terminal tracking loss, $\mathcal{L}_{\text{DPC}} = \|u(\cdot,T) - u_{\text{target}}\|^2$. The initial condition $u_0$ is sampled from a Gaussian random field, $\text{GRF}(l = 0.2, \sigma_u^2 = 4)$, 
and the target profiles $u_{\text{target}}$ are sampled from $\text{GRF}(l_T = 0.4, \sigma_{u,T}^2 = 4)$. The learned policy $\pi_{\mathbf{W}}(u(\cdot, t), u_{\text{target}}) \rightarrow \{f_i(t)\}_{i=1}^4$ maps current temperature distribution and target to control actions.

\noindent \textbf{Closed-Loop Results.} Figure~\ref{fig:heat1D_DPC_results} shows the closed-loop performance of the learned policy deployed on the high-fidelity FDM solver for two representative initial-condition/target pairs. The controlled trajectories converge to the prescribed targets with terminal errors of $\mathcal{O}(10^{-3})$, validating both the learned surrogate and the control policy. Notably, the control signals stabilize within the first few dozen time steps despite the absence of an explicit control-effort penalty in the training objective. We further benchmark the policy on $3000$ randomly sampled scenarios against nonlinear MPC (NMPC). As reported in Table~\ref{tab:table_comparison}, \D{which consolidates results for all three benchmark systems, the learned policy matches NMPC} in tracking accuracy while executing approximately four to five orders of magnitude faster per step, highlighting the computational benefits of offline policy amortization.

\subsection{Inviscid Burgers' Equation (BE): Shock Mitigation} 
\label{sec:burgers}

\begin{figure*}[!t]
    \centering
    \includegraphics[width=0.6\textwidth]{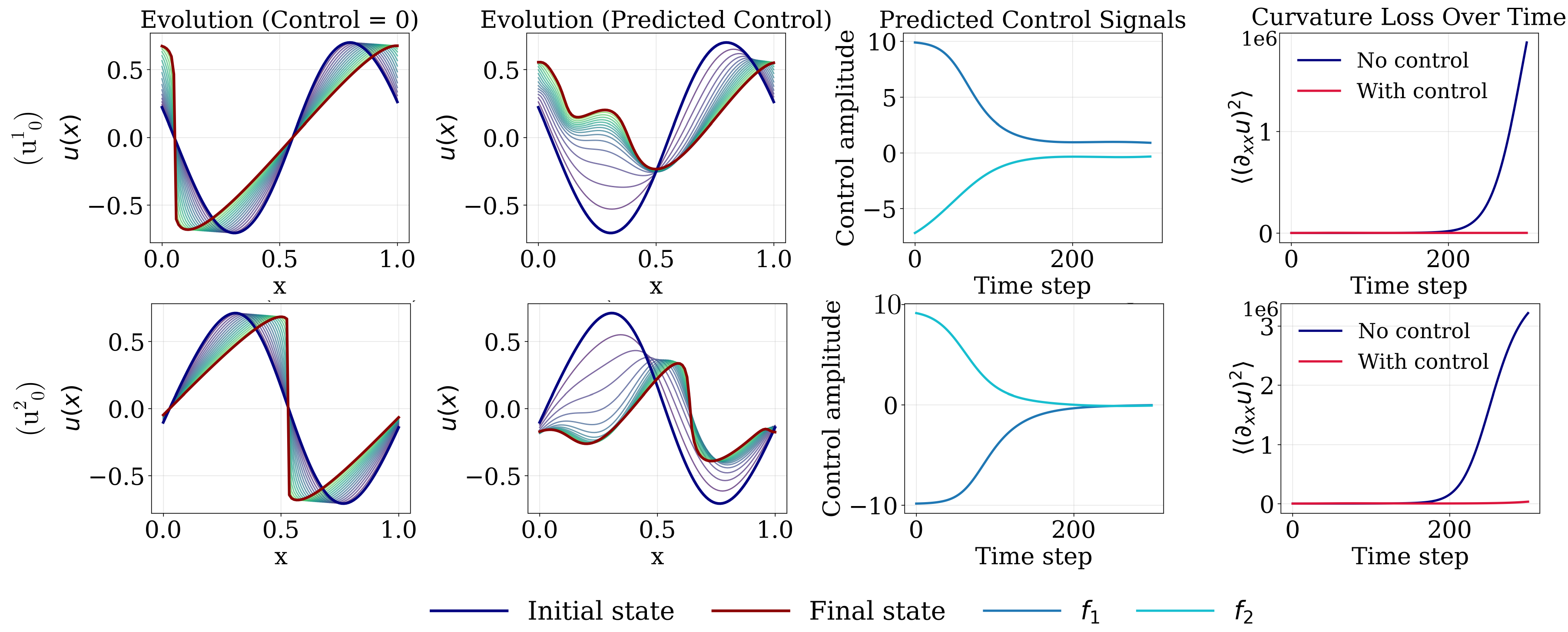}
    \vspace{-0.2cm}
    \caption{\D{BE shock control. Here $u_0^i$ denotes the initial state for the $i$-th scenario. Each row presents a scenario showing: \textit{(far left)} uncontrolled shock development; \textit{( middle left)} controlled smooth evolution; \textit{(middle right)} control signals $f_i(t)$; \textit{(far right)} curvature loss reduction.}}
    \label{fig:burger1D_DPC_results}
\end{figure*}

\textbf{System Dynamics.} Consider BE with periodic boundary conditions:
\begin{equation}
        \displaystyle\frac{\partial u}{\partial t} + u \frac{\partial u}{\partial x} = f(x,t), \quad u(0, t) = u(1,t),
\end{equation}
where $x \in [0, 1], \, t \in [0,T]$. $u(x,t)$ represents velocity and $T = 0.3$.  We deploy $n=2$ actuators at $\mu_i \in \{0.3, 0.6\}$ with $\sigma = 0.15$. This nonlinear hyperbolic PDE develops shock discontinuities, making it hard to control.

\noindent \textbf{Data Generation and Operator Learning.} FDM with first-order upwind scheme captures shock behavior. We generate 3000 training samples from GRF initial conditions (RBF kernel: $l = 0.25$, $\sigma_u^2 = 0.25$) with control bounds $|f_i(t)| \leq 10$. The trained TI-DeepONet achieves average relative $L_2$ error of $9.24 \times 10^{-2}$ on 3000 test samples.

\noindent \textbf{Control Objective.} Reduce shock formation by minimizing spatial curvature over time, $\mathcal{L}_{\text{DPC}} = \int_{0}^{T} \left\|\frac{\partial^2 u(\cdot,t)}{\partial x^2}\right\|_{L^2}^2 \, dt$. The initial condition $u_0$ is sampled from $\text{GRF}(l = 0.5, \sigma_u^2 = 0.25)$. The policy $\pi_{\mathbf{W}}(u(\cdot, t)) \rightarrow \{f_i(t)\}_{i=1}^2$ maps the current velocity profile to control amplitudes.

\noindent \textbf{Closed-Loop Results.} Figure~\ref{fig:burger1D_DPC_results} demonstrates that the learned policy achieves a 74.40\% reduction in curvature compared to uncontrolled evolution, effectively suppressing shock formation while respecting the imposed control constraints. As shown in the rightmost column, the curvature loss $(\partial_{xx} u)^2$ grows by several orders of magnitude under free evolution, whereas the controlled trajectories maintain near-zero curvature throughout the simulation horizon. Consistent with the heat equation results (Section~\ref{sec:heat}), the control signals converge to near-zero amplitude within the first few hundred time steps, indicating that the policy learns to stabilize the solution early and then withdraw actuation. We further evaluate the policy on $200$ randomly sampled initial conditions and compare against NMPC. The learned policy incurs approximately one order of magnitude higher loss than NMPC; however, it executes roughly five orders of magnitude faster per step (Table~\ref{tab:table_comparison}), presenting a favorable accuracy--speed trade-off for real-time deployment scenarios.

\subsection{Fisher-KPP Reaction-Diffusion Equation (RDE): Population Density Control} 
\label{sec:rd}

\textbf{System Dynamics.} Consider the RDE with Neumann (no-flux) boundaries:
\begin{equation}
        \displaystyle\frac{\partial u}{\partial t} = \alpha \frac{\partial^2 u}{\partial x^2} + ru(1-u) - f(x,t), ~~
        \displaystyle\frac{\partial u(0,t)}{\partial x} = \frac{\partial u(1,t)}{\partial x} = 0,
\end{equation}
where $x \in [0, 1], \, t \in [0,T]$. $u(x,t)$ represents population density, $\alpha=0.01$ is the diffusion coefficient, $r=1.0$ is the reaction rate, and $T = 0.3$. This system models population dynamics and chemical reactions. Control uses $n=4$ actuators at $\mu_i \in \{0.2, 0.4, 0.6, 0.8\}$ with $\sigma = 0.1$.

\noindent \textbf{Data Generation and Operator Learning.} FDM with backward Euler and Newton iteration handles the nonlinear reaction term. Training data comprises 3000 samples from GRF initial conditions ($l = 0.2$, $\sigma_u^2 = 0.25$) with control bounds $|f_i(t)| \leq 10$. The trained TI-DeepONet achieves average relative $L_2$ error of $1.18 \times 10^{-2}$ on test data.

\noindent \textbf{Control Objective.} Steer the density field to a desired target configuration, $\mathcal{L}_{\text{DPC}} = \|u(\cdot,T) - u_{\text{target}}\|^2$. The initial condition $u_0$ is sampled from a Gaussian random field, $\text{GRF}(l = 0.5, \sigma_u^2 = 0.25)$, 
and the target profiles $u_{\text{target}}$ are randomly selected from the training data.

\begin{figure}[!bt]
    \centering
\includegraphics[width=0.5\textwidth]{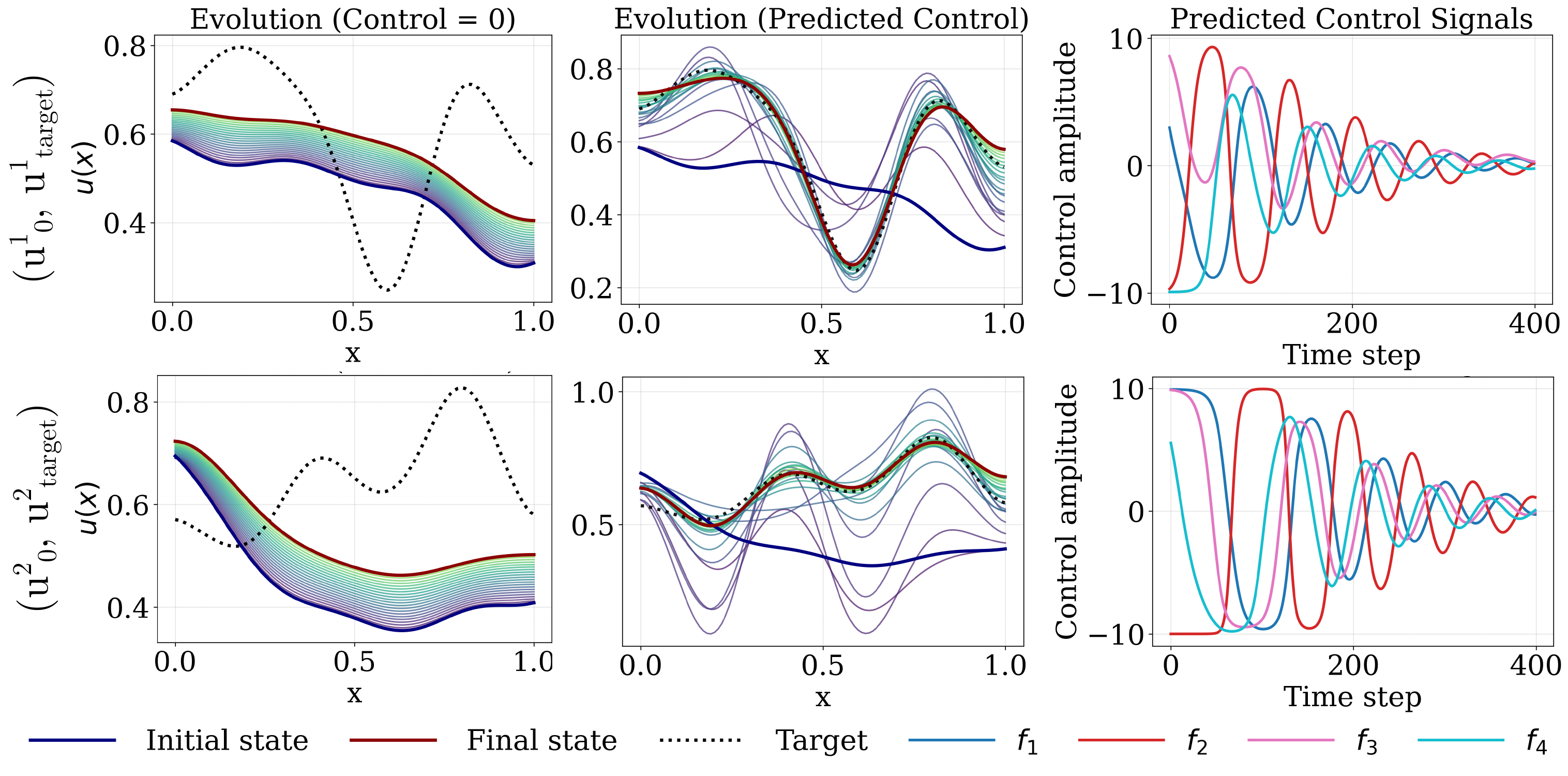}
        \vspace{-0.6cm}
    \caption{RDE density control. Here $u_0^i$ denotes the initial state and $u_{\mathrm{target}}^i$ the target state for the $i$-th scenario. Each row presents a scenario showing: \textit{(left)} uncontrolled evolution from initial state (blue) to final state (red) versus target (black dotted); \textit{(middle)} controlled trajectory achieving the target; \textit{(right)} applied control signals $f_i(t)$.}
    \label{fig:RD_1D_DPC_results}
\end{figure}

\noindent \textbf{Closed-Loop Results.} Figure~\ref{fig:RD_1D_DPC_results} shows successful density control with terminal error $\mathcal{O}(10^{-4})$ on the FDM solver, demonstrating effective handling of nonlinear reaction-diffusion dynamics. Unlike the heat and Burgers' equations, the control signals exhibit sustained oscillatory behavior throughout the simulation horizon. This is attributable to the stronger nonlinearity of the reaction term, which continually drives the solution away from the target and demands persistent corrective actuation. Despite this more demanding control landscape, the policy generalizes well: over $200$ randomly sampled initial-condition/target pairs, the learned policy achieves tracking loss of the same order of magnitude as NMPC while requiring approximately four orders of magnitude less computation time per step (Table~\ref{tab:table_comparison}). Among the three benchmarks, the reaction-diffusion equation thus represents the most challenging control task, yet the neural policy remains competitive with online optimization at a fraction of the computational cost.

\begin{table}[tb]
\D{
\caption{Performance Comparison of NMPC and TI-DeepONet DPC Algorithms}
\label{tab:table_comparison}
\centering
\resizebox{\columnwidth}{!}{%
\begin{tabular}{llccc}
\toprule
\textbf{Algorithm} & \textbf{Metric} & \textbf{HE ~(\ref{sec:heat}}) & \textbf{BE~(\ref{sec:burgers}}) & \textbf{RDE~(\ref{sec:rd}}) \\
\midrule
\multirow{2}{*}{NMPC}   & MSE           & $(4.2\pm6.7) \times 10^{-3}$ &$(3.1\pm2.9) \times 10^{+6}$ & $(1.8\pm1.6) \times 10^{-3}$ \\
                        & Time/step (s) & $4.3\pm1.5$                  & $3.9\pm13$   & $7.9\pm1.6$   \\
\addlinespace
\multirow{2}{*}{TI-DeepONet DPC} & MSE           & $(7.7\pm20) \times 10^{-3}$  & $(4.4\pm9.3) \times 10^{+7}$ & $(1.4\pm1.4) \times 10^{-3}$   \\
                        & Time/step (s)      & $(2.5\pm2.2) \times 10^{-4}$ & $(8.8\pm1.3) \times 10^{-5}$   & $(2.4\pm2.4) \times 10^{-4}$   \\
\bottomrule
\end{tabular}%
}}
\end{table}

\section{CONCLUSIONS AND FUTURE WORKS}


We presented a differentiable predictive control framework for PDE-constrained optimal control that integrates Time-Integrated Deep Operator Networks with offline policy optimization. By learning instantaneous temporal derivatives and employing  numerical integrators, our approach preserves temporal causality while enabling efficient gradient-based policy learning through automatic differentiation. Our experiments on three canonical PDEs (heat equation, Burgers' equation, and Fisher-KPP equation) demonstrate that the learned neural control policies successfully achieve diverse control objectives, including target tracking, shock mitigation, and population density regulation\D{, with accuracy comparable to nonlinear MPC while being four to five orders of magnitude faster at inference time}. The policies generalize effectively across distributions of initial conditions and problem parameters while maintaining constraint satisfaction. Notably, policies trained on the TI-DeepONet surrogate transfer successfully to high-fidelity finite difference solvers, validating the quality of the learned dynamics model.


Future directions include extending the framework to strongly nonlinear hyperbolic PDEs, scaling to two- and three-dimensional spatial domains with efficient surrogate architectures, \D{incorporating Lyapunov-based stability and probabilistic constraint-satisfaction guarantees to certify closed-loop performance for safety-critical applications, and analyzing policy robustness to surrogate model mismatch.}



\bibliographystyle{IEEEtran}
\bibliography{reference}

\end{document}